\documentclass[11pt]{article}

\usepackage{graphicx}
\usepackage{mathrsfs}
\usepackage{latexsym}
\usepackage{graphics}
\usepackage{hyperref}
\usepackage{epstopdf}
\pdfoutput=1

\begin{document}

\title{Hydrodynamic of wine swirling}

\author{M. Reclari$^1$, M. Dreyer$^1$, S. Tissot$^2$, D. Obreschkow$^1$, F. M. Wurm$^2$, M. Farhat$^1$ \\
\\ 1 EPFL, Laboratoire des Machines Hydrauliques, 1007 Lausanne, Switzerland\\2 EPFL, Laboratory of Cellular Biotechnology, 1015 Lausanne, Switzerland}

\maketitle

Various steps are involved with wine tasting, sometimes referred as the five S: See Swirl Smell Sip Savor. Of those step, the swirling is necessary to release the bouquet of the wine and is usually obtained by a gentle circular movement of the glass. The wave generated by this movement propagates along the glass walls and enhances the oxygenation and the mixing. We here define ``orbital shaking'' as the motion on a circular trajectory, at a constant angular velocity, of a cylindrical container maintaining a fixed orientation with respect to an inertial frame of reference. Recently, the orbital shaking has been applied to large scale bioreactors for cultivations of cells expressing recombinant proteins (e.g.~antibodies). A deeper understanding of the phenomena involved with the orbital shaking is thus required.

\begin{figure}[h!]
  \includegraphics[width=\columnwidth]{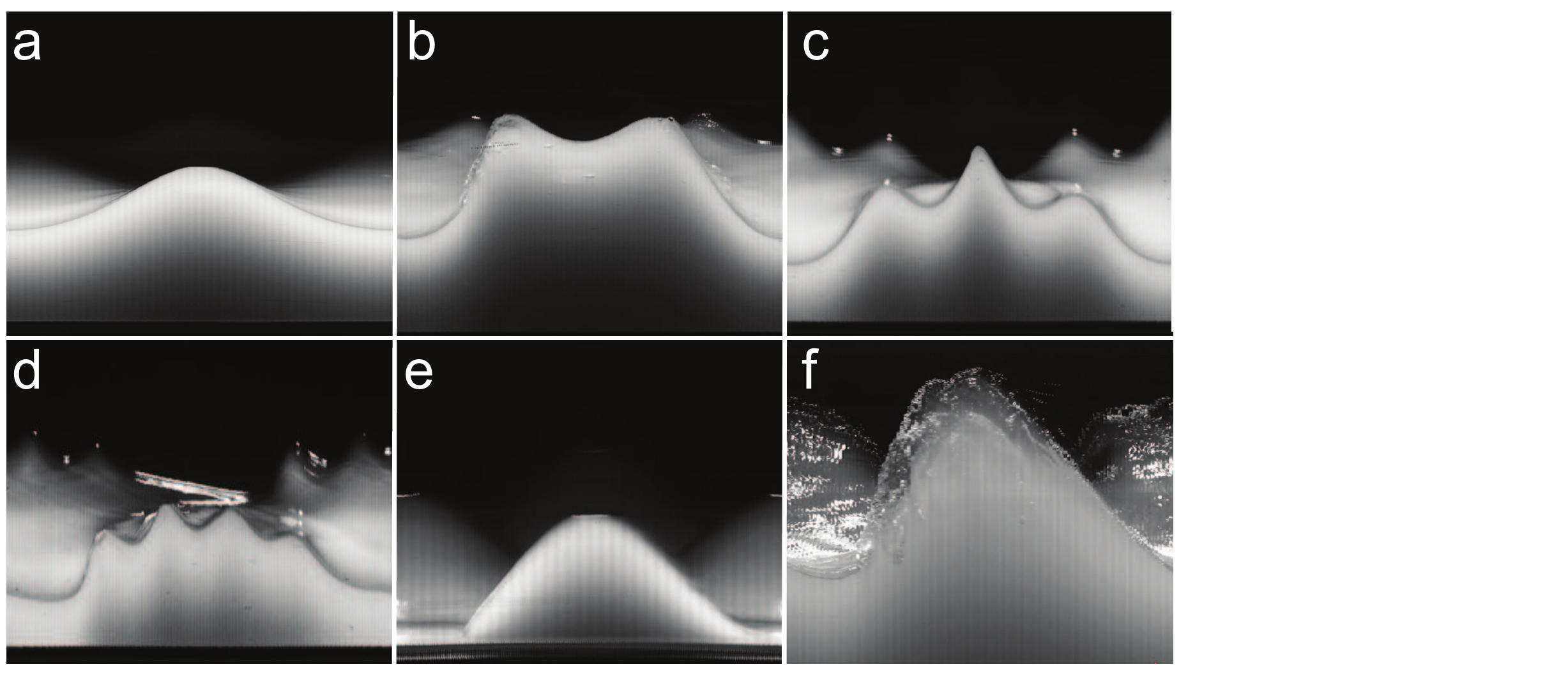}
  \caption{Wave patterns: (a) Single crest wave.  (b) Double crest wave. (c) Triple crest wave. (d) Quadruple crest wave. (e) Single crest wave drying a portion of the container bottom. (f) Breaking single crest wave. }
  \label{WavePatternFigure}
\end{figure}

We have investigated the shape of the wave generated by the orbital shaking using a circular cylinder with flat bottom. The free parameters are the inner diameter of the vessel $D$, the diameter of the shaking trajectory $d_s$, the elevation of the water at rest $H_0$ and the angular velocity $\omega$. The film shows a large variety of observed wave shapes: the most simple being a wave with one crest and one trough, as depicted in Fig.~\ref{WavePatternFigure}(a). More complex shapes, featuring multiple crests and troughs, are then shown: double (Fig.~\ref{WavePatternFigure}(b)), triple (Fig.~\ref{WavePatternFigure}(c)) and quadruple crest waves (Fig.~\ref{WavePatternFigure}(d)). Under certain conditions the wave could 'dry' a portion of the vessel bottom (Fig.~\ref{WavePatternFigure}(e)) or break (Fig.~\ref{WavePatternFigure}(f)). 

Performing a dimensional analyses taking into account also the gravitational acceleration, we have identified three dimensionless parameters governing the shape of the free surface: $\tilde{\delta}\equiv\delta/D$, $\tilde{d}_s\equiv d_s/D$, $\tilde{H}_0\equiv H_0/D$ and $Fr^2\equiv(\omega^2 d_s)/g$. The video shows the equivalence between the wave shapes obtained with the same dimensionless parameters but at overall different scales. Each combination of the free parameter has a peculiar balance of forces, generating a particular wave shape.

\end{document}